\documentclass[onecolumn]{aa} 

\usepackage{graphicx}
\usepackage{txfonts}

\begin{document}

   \title{Test magnetohydrostatic extrapolation with\\ radiative MHD simulation of a solar flare}

   \author{X. Zhu
          \inst{1}
          \and
          T. Wiegelmann\inst{1}
          }

   \institute{Max-Planck-Institut f\"{u}r Sonnensystemforschung, Justus-von-Liebig-Weg 3, 37077 G\"{o}ttingen, Germany\\
              \email{zhu@mps.mpg.de}
             }

   \date{Received ; accepted }

 
  \abstract
   {On the sun, the magnetic field vector is measured routinely only in the photosphere. By using these photospheric measurements as boundary condition, we developed the magnetohydrostatic (MHS) extrapolation to model the solar atmosphere. The model makes assumption about the relative importance of magnetic and non-magnetic forces. While the solar corona is force-free, this is not the case in photosphere and chromosphere.}
   {The model has been tested with an exact equilibria in \cite{zw18}. Here we present a more challenging and realistic test of our model with radiative MHD simulation of a solar flare.}
   {By using the optimization method, the MHS model computes self-consistently the magnetic field, plasma pressure and density. The nonlinear force-free field (NLFFF) and gravity stratified atmosphere along the field line are assumed as the initial condition of the optimization.}
   {Compared with NLFFF, the MHS model gives an improved magnetic field not only in magnitude and direction, but also in the magnetic connectivity. Besides, the MHS model is able to recover the main structure of the plasma in the photosphere and chromosphere.}
   {}


\titlerunning{MHS extrapolation validation}
   \maketitle


\section{Introduction}
State of the art model of the solar coronal magnetic field is the so-call nonlinear force-free field (NLFFF; cf. Wiegelmann \& Sakurai \citeyear{ws12} and Guo et al. \citeyear{gcd17}) because of the low plasma $\beta$ \citep{g01} in the corona. This, however, is not the case in the photosphere and chromosphere where the plasma $\beta$ is close to or even larger than unity. By calculating the net Lorentz force of AR7216, \cite{mjm95} found that the magnetic field is not force-free in the photosphere. \cite{zwd16,zwc17} derived the magnetic field configuration of active regions using the MHD relaxation approach. They found the force-free assumption failed beneath the height 1.8 Mm. To study the magnetic field in the non-force-free region, a magnetohydrostatic (MHS) extrapolation should be used.

The MHS extrapolation is introduced based on the MHS equilibria assumption which is supposed to be a better approximation in the low layers of the Sun than force-free assumption. The governing equations of the MHS equilibria can be written as
\begin{eqnarray}
(\nabla \times \mathbf{B})\times \mathbf{B}-\nabla p - \rho \mathbf{\hat{z}} & = & 0, \label{eq:force_balance}\\
\nabla \cdot \mathbf{B} & = & 0,
\end{eqnarray}
where $\mathbf{B}$, $p$ and $\rho$ are the magnetic field, plasma pressure and plasma density, respectively. Notice that the equations above have been normalized using the following constants: $\rho_0=2.7\times10^{-1}g/cm^3$ (density), $T_0=6\times 10^3K$ (temperature), $g=2.7\times10^4 cm/s^2$ (gravitational acceleration), $L_0=\frac{\mathcal{R}T_0}{\mu g}=1.8\times10^7 cm$ (length), $p_0=\sqrt{\frac{\rho_0\mathcal{R}T_0}{\mu}}=1.3\times10^5 dyn/cm^2$ (plasma pressure), and $B_0=\sqrt{4\pi p_0}=1.3\times10^3 G$ (magnetic field). \cite{l85,l91,l92} and \cite{nr99} found the MHS equations can be linearized when the current consists of a linear component parallel to the magnetic field and another component perpendicular to gravity. Using this so-called linear MHS model, \cite{ads98,adm99} derived the magnetic field and plasma distribution of AR7722 and AR7986. Recently, \cite{wnn15,wnn17} used this model to extrapolate the MHS equilibria by SUNRISE/IMaX magnetogram. The unprecedented resolution ($40\ km/pixel$) of IMaX enabled the model to resolve the thin non-force-free layer with several tens of grids. However, the linear MHS model excludes the strong concentration of electric currents and Lorentz forces. This is somewhat similar to the well known limitations of linear force-free field to model the corona.

A few methods have been developed to solve the MHS equations in the general case numerically. \cite{hd06,hd08} proposed an approach to derive the non-force-free field by superposing one potential field and two linear force-free fields. The Grad-Rubin iteration procedure has been extended by \cite{gw13} and \cite{gbb16} to compute the MHS equilibria. The MHD relaxation method which applies ``evolution technique'' is able to yield the MHS solution \citep{mm94,jmm97,zwd13}. \cite{wi03}, \cite{wn06} and \cite{wnr07} used the optimization method to treat the MHS equations without gravity in different coordinates systems. Recently, we extended the optimization method to model the system with a gravity force in \cite{zw18} (hereafter Paper I). In that work, we tested our model with a perfect MHS equilibria \citep{l85,l91}. However, the real Sun is really dynamic and a lot more complex. We aim to investigate here if the MHS extrapolation works in a realistic situation.

In this paper, we apply our numerical code to reconstruct a snapshot of the solar flare simulation. The simulation includes 3D radiative transfer in the convection zone and photosphere, optically thin radiation and field aligned heat conduction in the corona, etc. These features make the simulation a pretty realistic one. The organization of the paper is as follows. In Sect.~\ref{sec:reference model} we describe briefly the simulation and assess the physical state of the reference snapshot. In Sect.~\ref{sec:method} we introduce the optimization method for the MHS model and the numerical setup to extrapolate the reference snapshot. In Sect.~\ref{sec:result} we evaluate the results through several metrics. In Sect.~\ref{sec:discussion}, we discuss a few factors that can impact the ability of our model to reconstruct the magnetic field and plasma. Concluding remarks are presented in Sect.~\ref{sec:conclusion}.


\section{The active region in the radiative MHD simulation}\label{sec:reference model}

Recently, \citet{crc19} carried out a radiative MHD simulation of a solar flare. They used the $\mathbf{M}$ax-Planck-Institute for Aeronomy/ $\mathbf{U}$niversity of Chicago $\mathbf{Ra}$diation $\mathbf{M}$agneto-hydrodynamics (MURaM) code which allows for simulations spanning from the upper convection zone into the solar corona \citep{vss05,r17}. The initial setup consists of a bipolar sunspots. After the thermally relaxed equilibra are reached, they impose the emergence of another twisted bipolar flux system at the bottom boundary. The accumulation of the magnetic energy finally leads to the abrupt relaxation, which powers coronal mass ejection and a flare.

\begin{figure}[b]
\centering
\includegraphics[width=\hsize*3/4]{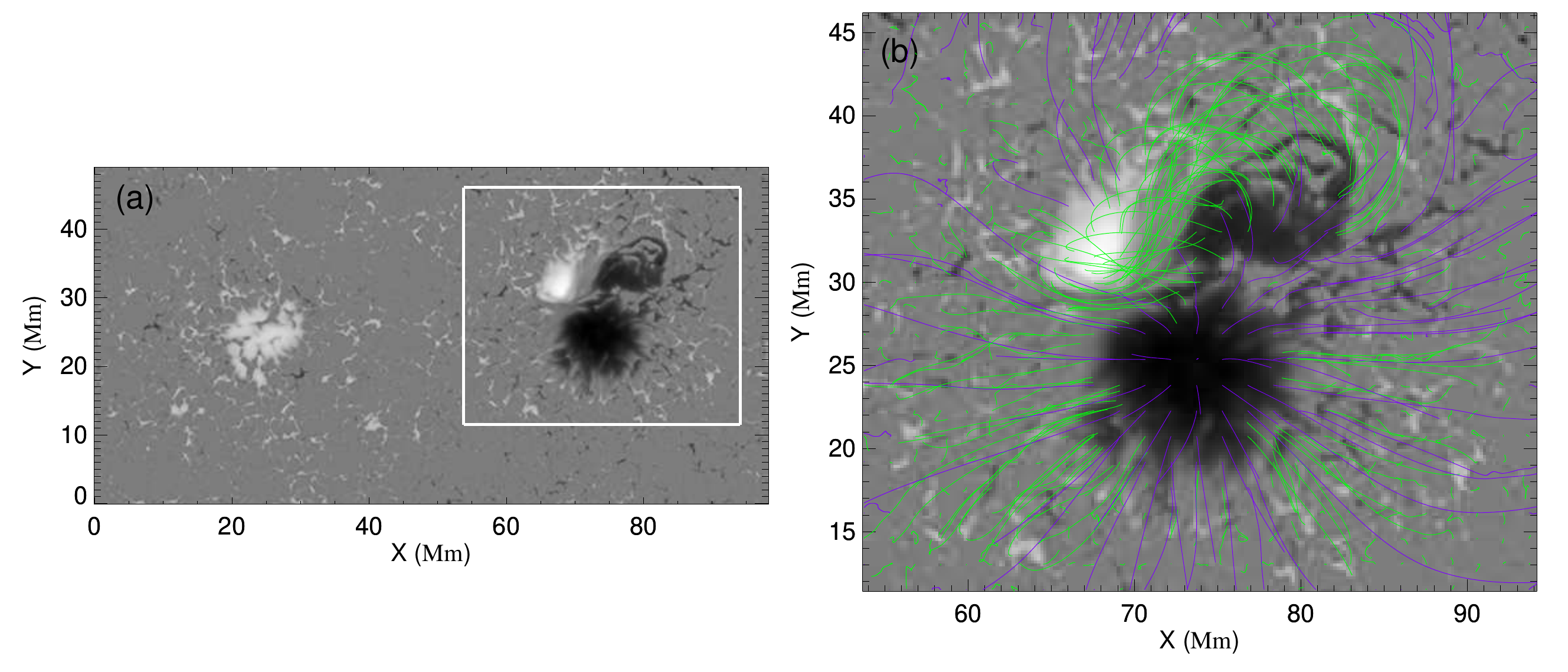}
\caption{(a) Magnetogram in the height where $<\!\!\tau\!\!>=1$, which is used as the boundary input. The ``hgcr\_hslice.pro'' from the data link is used to derive the MHD quantities sampled at constant height. For photosphere where $<\tau>=1$, set input parameter k=116. (b) Magnetic field line patterns in the box outlined in panel (a). Purple/green lines represent open/closed field lines.}
\label{fig:magnetogram_linepattern}
\end{figure}

Among all the snapshots that are available for download (http://purl.stanford.edu/dv883vb9686), we choose the last one (about 8 minutes after the flare) as the reference model used for the test. The original data spans $\pm 49.152$ and $\pm 24,576$ Mm in the x- and y-axes, respectively. In the z-axis, the data spans from 7.5 Mm beneath the photosphere to 41.6 Mm above. The grid spacing is 192 (64) km in the horizontal (vertical) direction. In this work, we focus on the MHS equilibrium in the lower atmosphere. For this purpose, the data from photosphere to 8.192 Mm above are extracted. The photosphere cuts at the average geometrical height corresponding to optical depth unity. This is done by using ``hgcr\_hslice.pro'' with input parameter ``k = 116'' which is suggested in ``readme.txt''. Both ``hgcr\_hslice.pro'' and ``readme.txt'' can be downloaded from the above data link. Photospheric magnetogram and the magnetic field lines in the sub-volume are illustrated in Fig.~\ref{fig:magnetogram_linepattern}.

\begin{figure}[t]
\centering
\includegraphics[width=\hsize*3/4]{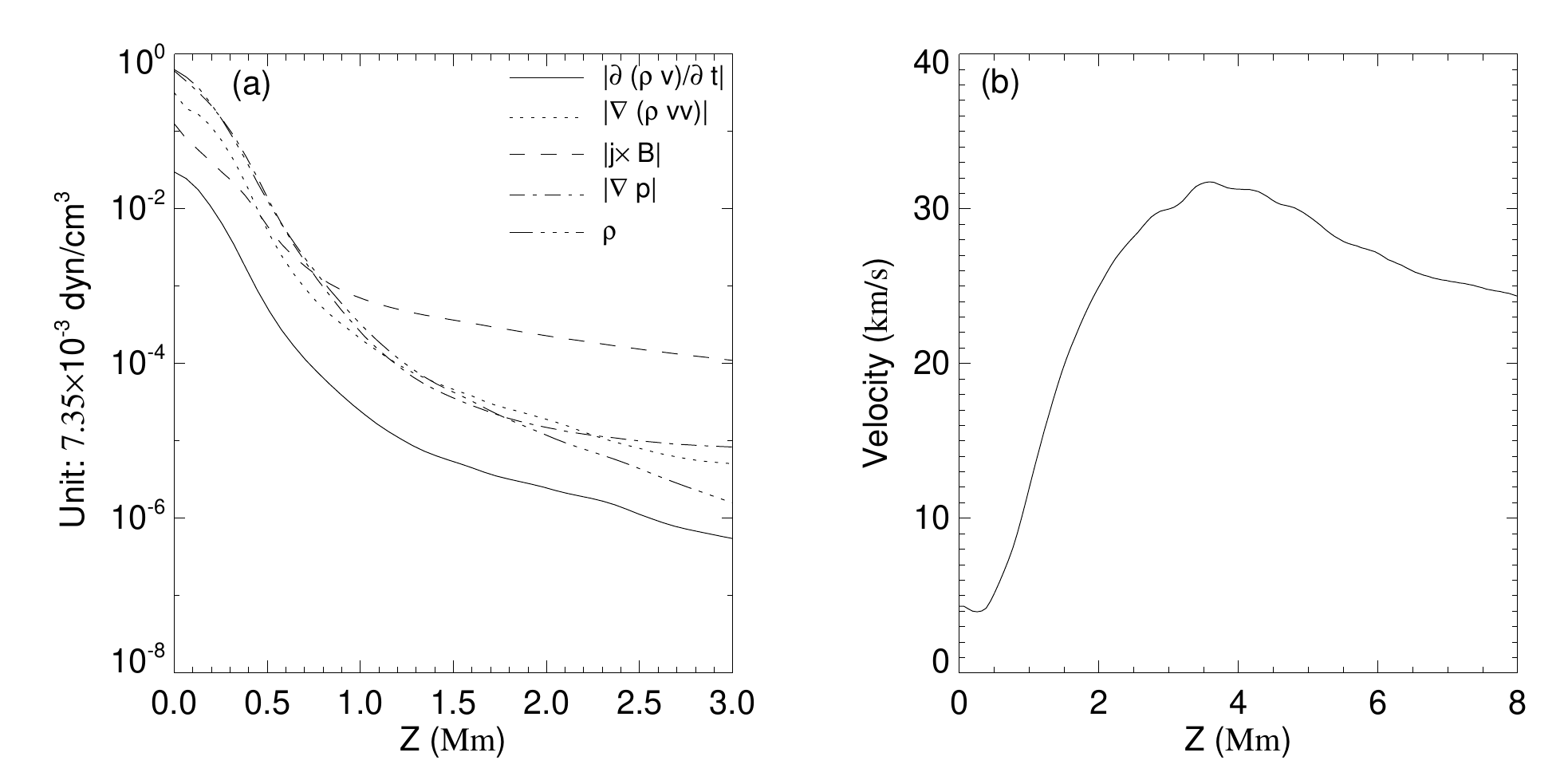}
\caption{(a) Planar averaged value of each term in the momentum equation. (b) Planar averaged velocity varies along height.}
\label{fig:time_varying_term}
\end{figure}

\begin{figure}[t]
\centering
\includegraphics[width=\hsize*3/4]{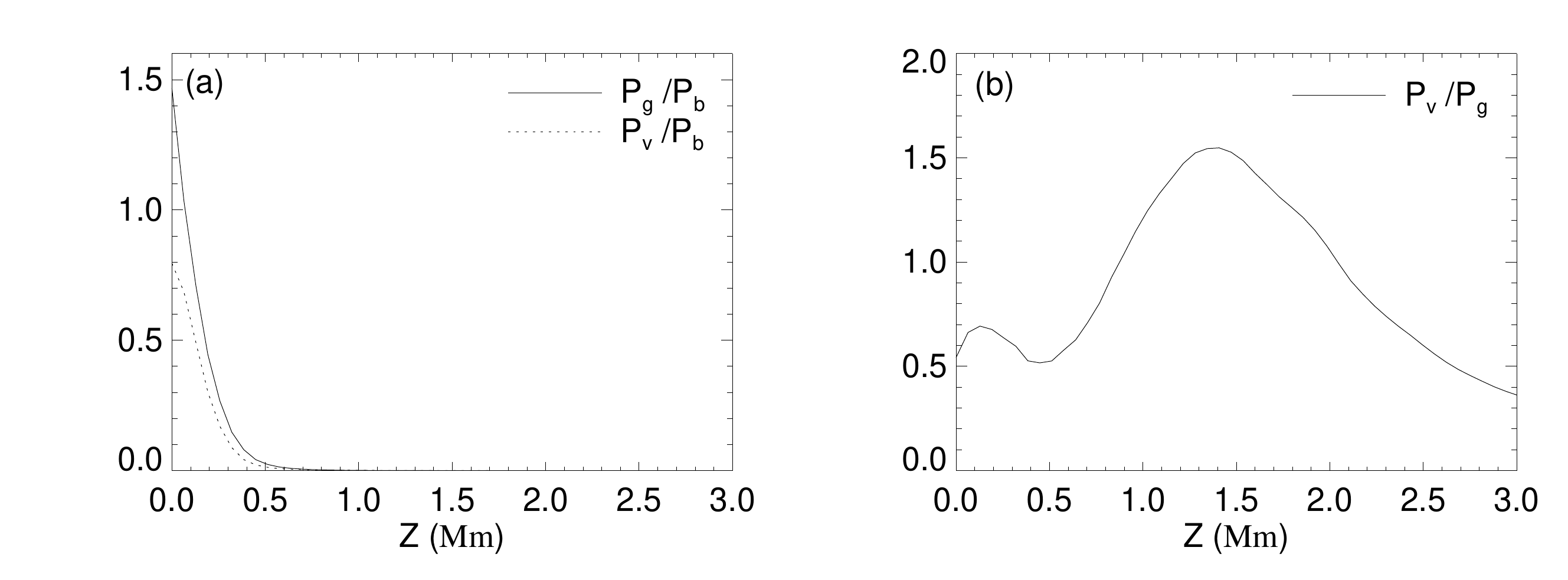}
\caption{(a) Horizontal averages for the reference model in the sub-volume outlined in Fig~.\ref{fig:magnetogram_linepattern} (a). The inertial term $\frac{1}{N}\sum \rho v^{2}$ and plasma pressure $\frac{1}{N}\sum p$ are plotted relative to the horizontally averaged magnetic pressure $\frac{1}{N}\sum B^{2}/2$. (b) the inertial term $\frac{1}{N}\sum \rho v^{2}$ is plotted relative to the horizontally averaged plasma pressure $\frac{1}{N}\sum p$.}
\label{fig:PbPgPv}
\end{figure}

The system develops all the time during the simulation. However, Fig.~\ref{fig:time_varying_term} (a) shows, at the time we concern, the time-varying term is smaller than other terms in the momentum equation. The reference snapshot is not strictly static (see panel b). 
Part of the velocity come from the violence of the flare eruption. To see how the moving fluid affect the magnetic field and plasma, we look at the momentum equation in steady state.
\begin{eqnarray}
  \nabla\cdot\left[\rho \mathbf{v}\mathbf{v}-\mathbf{B}\mathbf{B}+\left(\frac{B^{2}}{2}+p\right)\mathbf{I}\right] =0,
\end{eqnarray}
where $\rho$, $p$, $\mathbf{v}$ and $\mathbf{B}$ are mass density, pressure, velocity and magnetic field, respectively.  The inertial term ($P_{v}=\rho v^{2}$), plasma pressure ($P_{g}=p$) and magnetic pressure($P_{b}=B^{2}/2$) in the above equation compensate with each other. Fig.~\ref{fig:PbPgPv} shows the influence of these three terms. We see from panel (a) that: (1) plasma pressure affect the magnetic field significantly near the photosphere, (2) magnetic field dominate the three terms above 300 km. Panel (b) shows that the strength of inertial term is stronger than that of plasma pressure in the layer $1 Mm <z< 2 Mm$. We deduce that, in the extrapolation, (1) plasma pressure should be taken into account near the photosphere, (2) excluding the inertial term has limited impact on the pattern of magnetic field, (3) excluding the inertial term has large impact on the plasma distribution above 1 Mm. We conclude that this snapshot is suitable, not perfect though, to test the MHS extrapolation.

\section{The magnetohydrostatic extrapolation}\label{sec:method}

\subsection{Method}
The optimization method used in this work to solve the MHS equations was described in detail in Paper I. Here we give a short outline and its recent improvements. The method involves the minimization of the functional 
\begin{equation}
\begin{aligned}
L(\mathbf{B},p,\rho)&=\int_{V}\omega_{a}B^{2}\Omega_{a}^{2}+\omega_{b}B^{2}\Omega_{b}^{2}dV\\
&+\nu \int_{S}([\mathbf{B}_{\!o\;\!\!p\;\!\!t}\ p_{\!o\;\!\!p\;\!\!t}\ \rho_{\!o\;\!\!p\;\!\!t}]-[\mathbf{B}_{\!o\;\!\!b\;\!\!s}\ p_{\!o\;\!\!b\;\!\!s}\ \rho_{\!o\;\!\!b\;\!\!s}])\cdot \\
&\ \ \ \ \ \mathbf{W}\cdot ([\mathbf{B}_{\!o\;\!\!p\;\!\!t}\ p_{\!o\;\!\!p\;\!\!t}\ \rho_{\!o\;\!\!p\;\!\!t}]-[\mathbf{B}_{\!o\;\!\!b\;\!\!s}\ p_{\!o\;\!\!b\;\!\!s}\ \rho_{\!o\;\!\!b\;\!\!s}])d^{2}S,
\end{aligned}
\label{eq:L}
\end{equation}
with
\begin{eqnarray}
\mathbf{\Omega_{a}} &=& \left[(\nabla \times \mathbf{B})\times \mathbf{B}-\nabla p - \rho \mathbf{\hat{z}}\right]/(B^2+p), \label{eq:Omga}\\
\mathbf{\Omega_{b}} &=& [(\nabla \cdot \mathbf{B})\mathbf{B}]/(B^2+p),\label{eq:Omgb}
\end{eqnarray}
where $\omega_{a}$ and $\omega_{b}$ the weighting functions and $\nu$ the Lagrangian multiplier. $\mathbf{W}$ is a diagonal matrix to incorporate the measurement error of the magnetic field and gas pressure. $[\mathbf{B}_{\!o\;\!\!p\;\!\!t}\ p_{\!o\;\!\!p\;\!\!t}\ \rho_{\!o\;\!\!p\;\!\!t}]$ and $[\mathbf{B}_{\!o\;\!\!b\;\!\!s}\ p_{\!o\;\!\!b\;\!\!s}\ \rho_{\!o\;\!\!b\;\!\!s}]$ are the optimized and observed quantities in the photosphere respectively. The weighting functions and slow boundary injection are found to be able to improve the results \citep{w04,wti12}. Here for simplicity, we set the weighting function to unity over the whole region and use a fixed lower boundary. Note that, in Eqs.~\ref{eq:Omga} and \ref{eq:Omgb}, the denominator ($B^2+p$) is different from that of the previous code ($B^2$). We make the change in order to avoid nonphysically large contribution to the functional $L$ from the weak magnetic field region. Then, the solution of the MHS equations can be optimized by
\begin{equation}
minimize \quad L(\mathbf{B},Q,R)
\end{equation}
with $p=Q^{2}$ and $\rho=R^{2}$ which ensure the positive $p$ and $\rho$.

The gradient descent method is used to find the minimum:
\begin{eqnarray}
\mathbf{B}_{n+1}=\mathbf{B}_{n}-\mu_{B}*\delta L_\mathbf{B},\label{eq:Bn+1} \\
Q_{n+1}=Q_{n}-\mu_{Q}*\delta L_Q,\label{eq:Qn+1} \\
R_{n+1}=R_{n}-\mu_{R}*\delta L_R,\label{eq:Rn+1}
\end{eqnarray}
where $\delta L_\mathbf{B}$, $\delta L_Q$ and $\delta L_R$ (see Appendix \ref{sec:variables}) are the functional derivatives respect to function $\mathbf{B}$, $Q$ and $R$, respectively. $\mu_B$, $\mu_Q$ and $\mu_R$ control the step length.

In Paper I, a pressure distribution on the photospheric pressure was derived by solving the following Poisson's equation
\begin{equation}
\Delta_{ph} p=\nabla\cdot\mathbf{f}_{ph}, \label{eq:poisson}
\end{equation}
where $\mathbf{f}_{ph}$ is the 2D horizontal Lorentz force on the photosphere, $\nabla_{ph}=\mathbf{\hat{x}}\partial_{x}+\mathbf{\hat{y}}\partial_{y}$, $\Delta_{ph}=\partial_{x}^{2}+\partial_{y}^{2}$ is the 2D Laplacian. In this study, however, the Poisson method does not work effectively because of the dynamics in the simulation. Instead, a simple vertical magnetic flux tube is assumed to derive the pressure distribution from the force balance condition. That is:
\begin{equation}
p+B_{z}^2/2=p_{quiet},
\label{eq:fluxtube}
\end{equation}
where $p_{quiet}$ is the plasma pressure in the quiet sun where the magnet field is extremely weak. We set $p=10^{-3}p_{quiet}$ when the negative pressure appeared from Eq.~\ref{eq:fluxtube}.

\begin{figure}[t]
\centering
\includegraphics[width=\hsize/2]{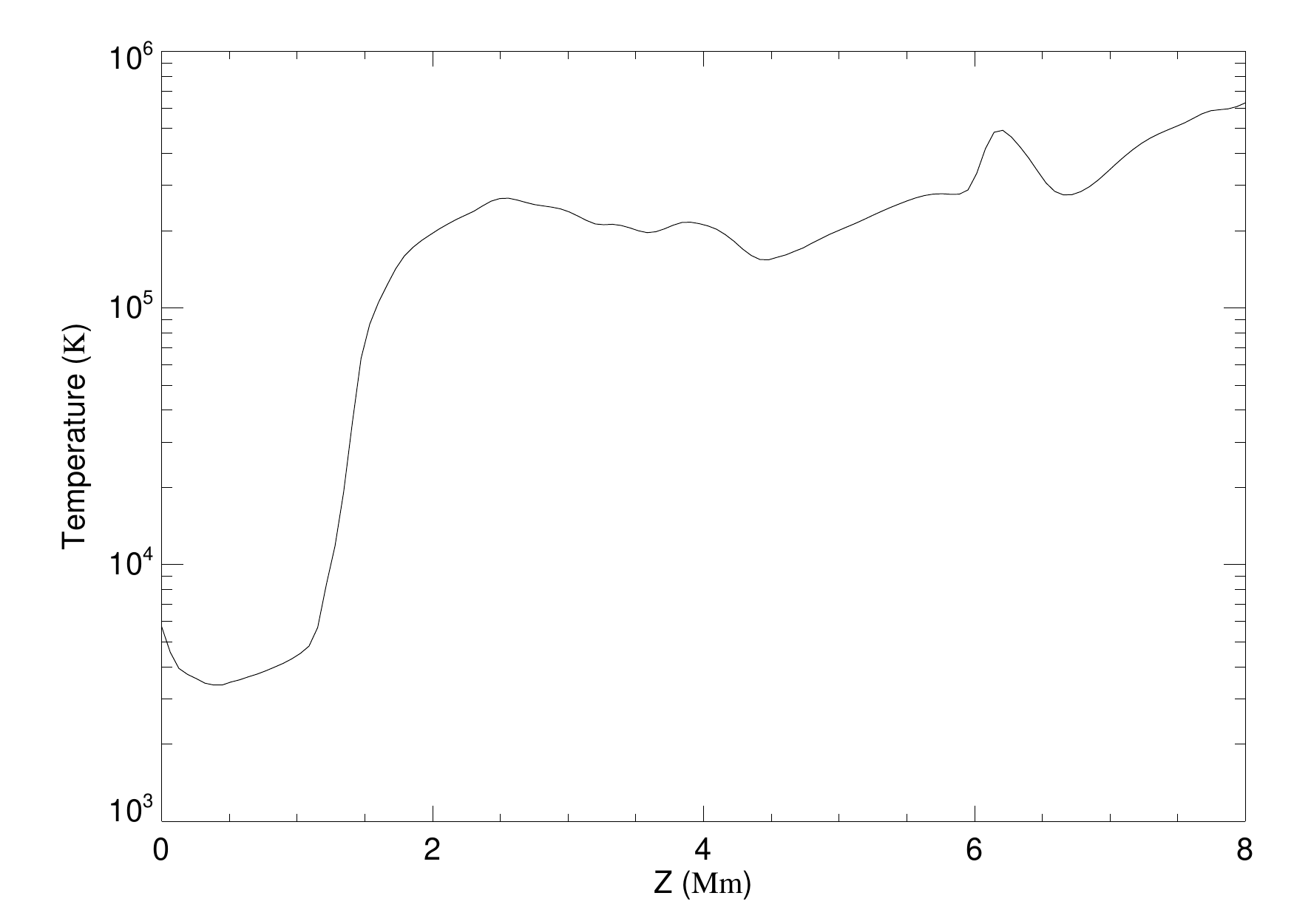}
\caption{Temperature profile of the equilibrium atmosphere in the initial condition.}
\label{fig:temperature}
\end{figure}

\subsection{Apply to a MURaM snapshot}
Base on a photospheric vector magnetogram from the MURaM simulation, we extrapolate the magnetic field, plasma pressure and density on $512\times256\times128$ grid points. The grid spacing (192/64km in horizontal/vertical direction) is the same with that in the simulation. The procedure shown below to perform the MHS extrapolation is slightly different from that was presented in Paper I:
\begin{enumerate}
  \item Calculate a NLFFF model \citep{w04,wis06}.
  \item Distribute the pressure in the photosphere by using $p+B_{z}^2/2=p_{quiet}$. Calculate the pressure along the magnetic field line with the gravity stratified assumption (see 1D temperature profile in Fig.~\ref{fig:temperature}). Calculate the density in the computational box by using ideal gas with a 1D temperature model.
  \item Iterate for $\mathbf{B}, Q$ and $R$ by Eqs.~\ref{eq:Bn+1}-\ref{eq:Rn+1}. This step is repeated until $L$ reaches its minimum.
\end{enumerate}

\section{Result}\label{sec:result}
\subsection{Plasma solution}
\begin{figure}[t]
\centering
\includegraphics[width=\hsize*11/23]{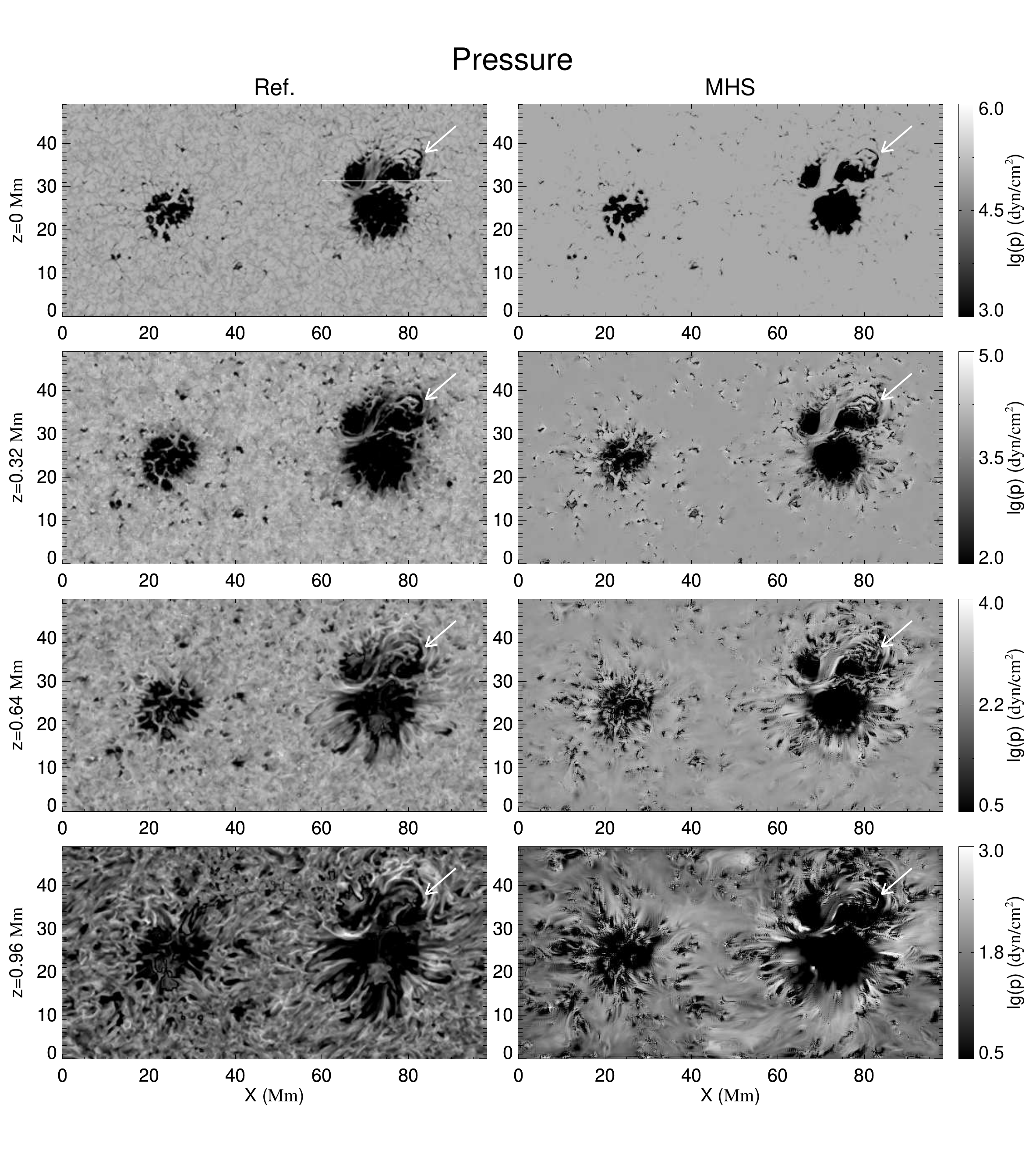}
\caption{Pressure distribution of reference (left) and MHS model (right) in different planes.}
\label{fig:p_horizontal}
\end{figure}

\begin{figure}[t]
\centering
\includegraphics[width=\hsize*11/23]{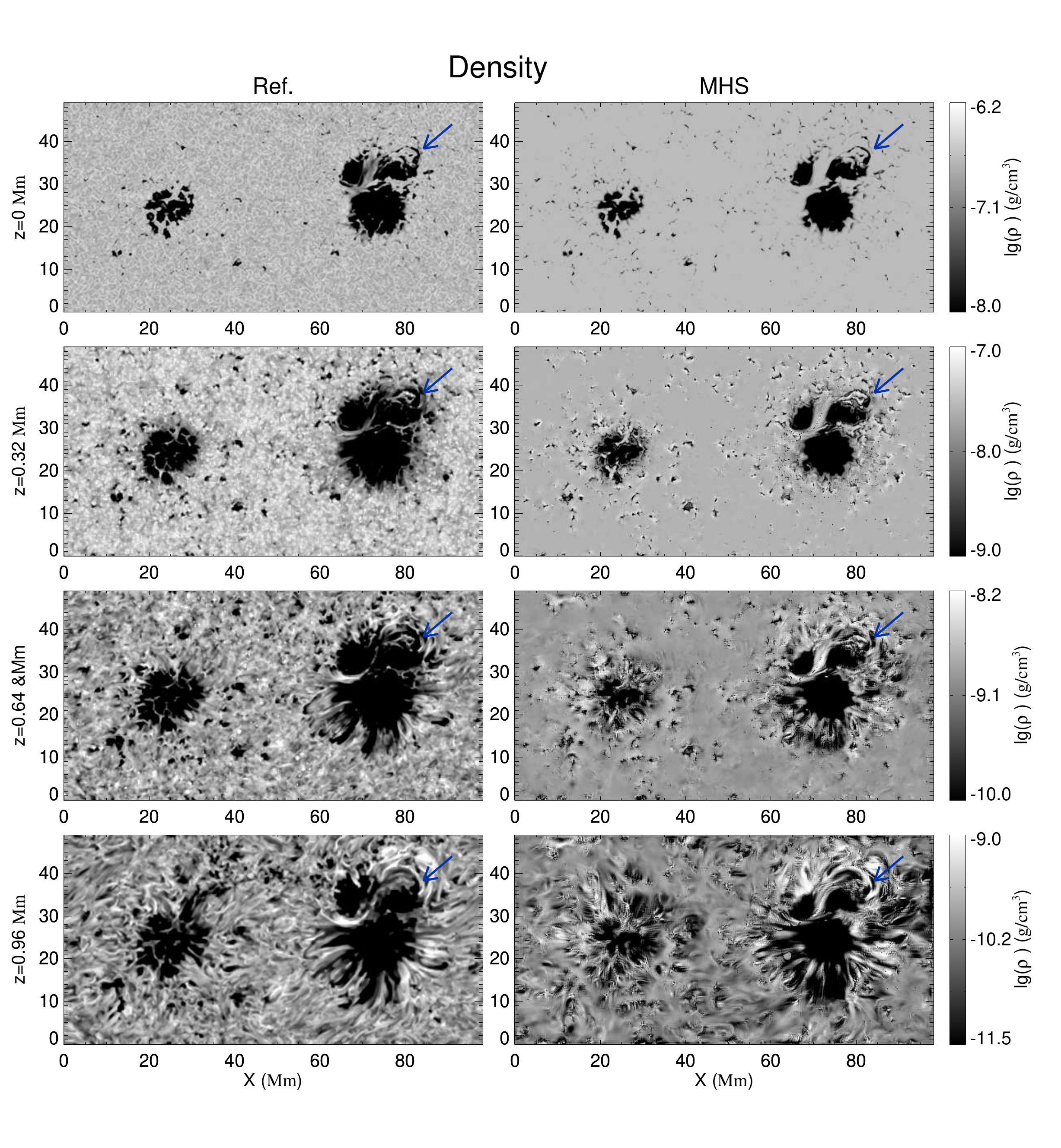}
\caption{Density distribution of reference (left) and MHS model (right) in different planes.}
\label{fig:d_horizontal}
\end{figure}

\begin{figure}[t]
\centering
\includegraphics[width=\hsize*2/3]{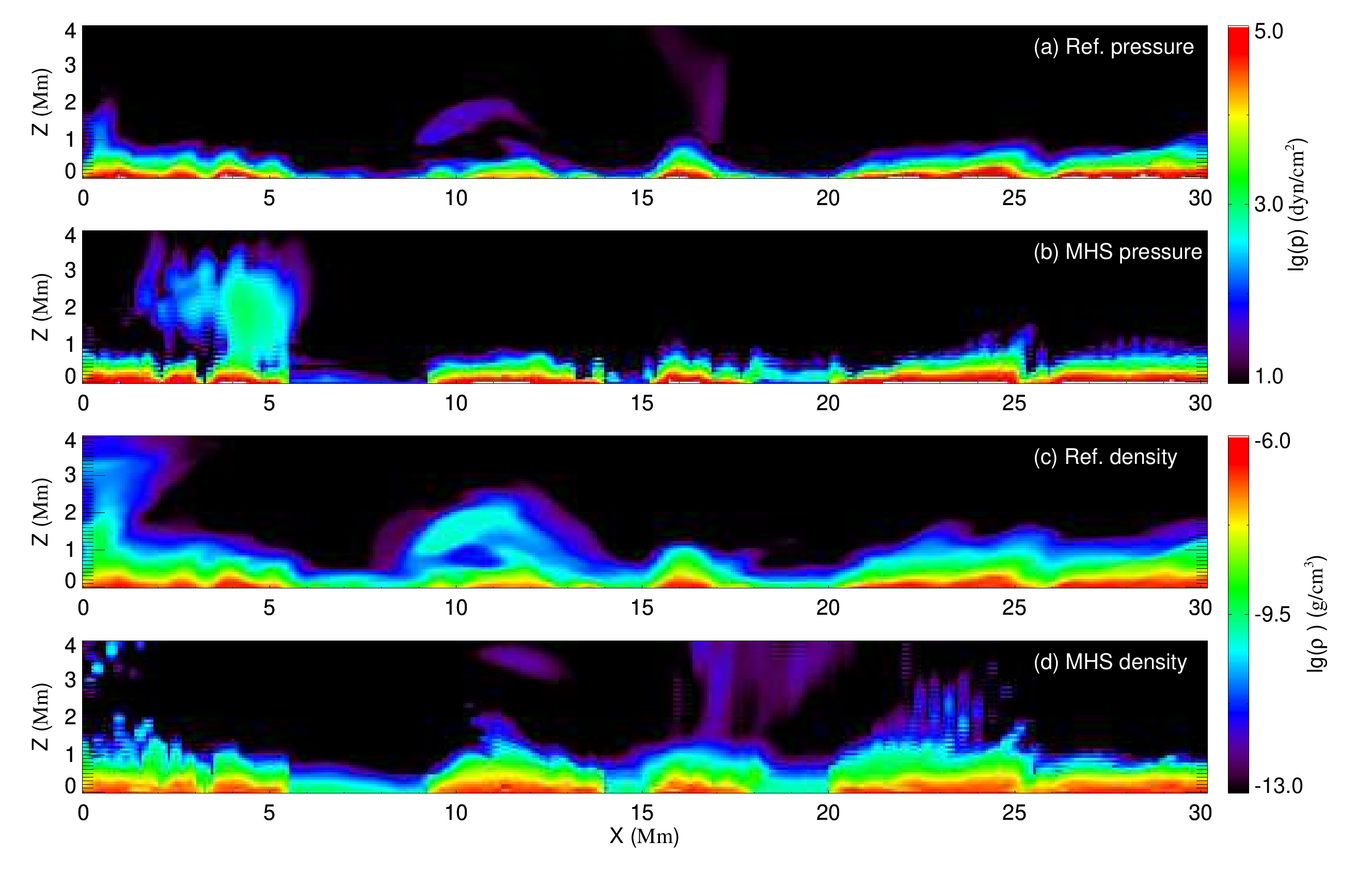}
\caption{Vertical slice along the cut indicated in Fig.~\ref{fig:p_horizontal}. (a)/(b) Pressure from reference/MHS model. (c)/(d) Density from reference/MHS model.}
\label{fig:pd_vertical}
\end{figure}

The MHS model computes the plasma in the computational box, which is an important advantage over NLFFF model. Fig.~\ref{fig:p_horizontal} and \ref{fig:d_horizontal} compare the pressure and density of the two models at different levels. In the bottom boundary, the plasma of the MHS model is highly similar with that of the reference model, which denotes $p+B_{z}^2/2=p_{quiet}$ does make sense. Above the bottom boundary, the MHS model is able to recover the main structures of the plasma in the phtosphere and lower chromosphere (below 1 Mm). For example: the depletion of pressure and density in the strong field region, the spiral structure (pointed by arrows in Fig.~\ref{fig:p_horizontal} and \ref{fig:d_horizontal}) in the new emerging region and the ``tentacle'' around the spots as well. From another perspective, Fig.~\ref{fig:pd_vertical} illustrates pressure and density in a vertical plane. We see clearly the main structures below 1 Mm are well recovered. Above 1 Mm, however, large deviation appears. As we did not use the temperature data in the modeling, the plasma can not be determined exactly. Another reason for the inaccurate plasma result is the dynamics of the atmosphere.

\begin{figure}[b]
\centering
\includegraphics[width=\hsize*2/3]{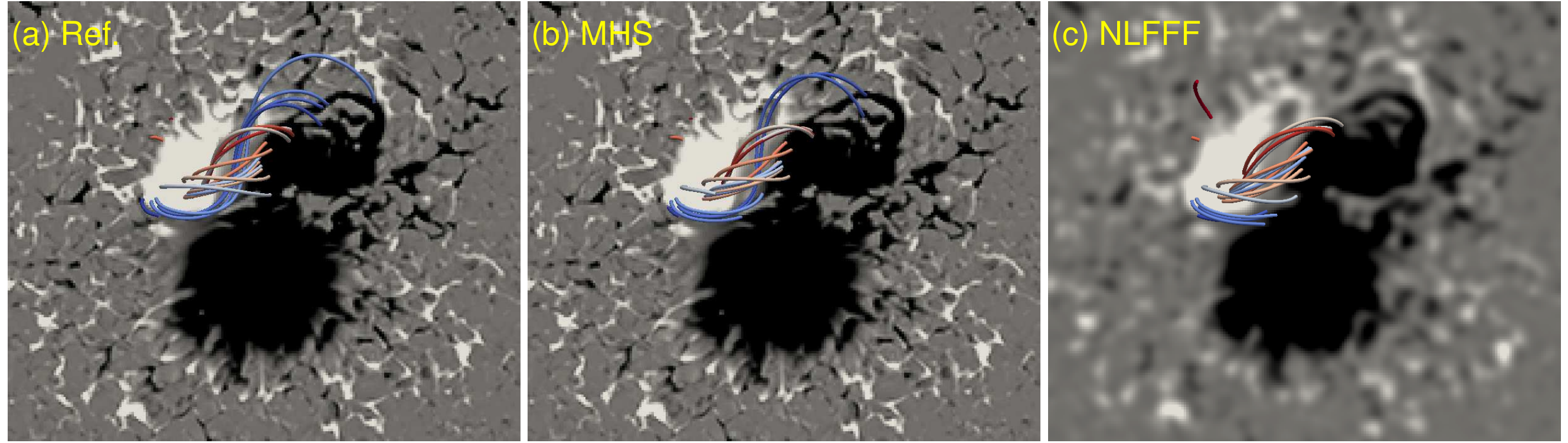}
\caption{Selected field lines of the magnetic flux tube.}
\label{fig:linecompare}
\end{figure}

\subsection{3D magnetic structure in the models}

\begin{figure}[t]
\centering
\includegraphics[width=\hsize*2/3]{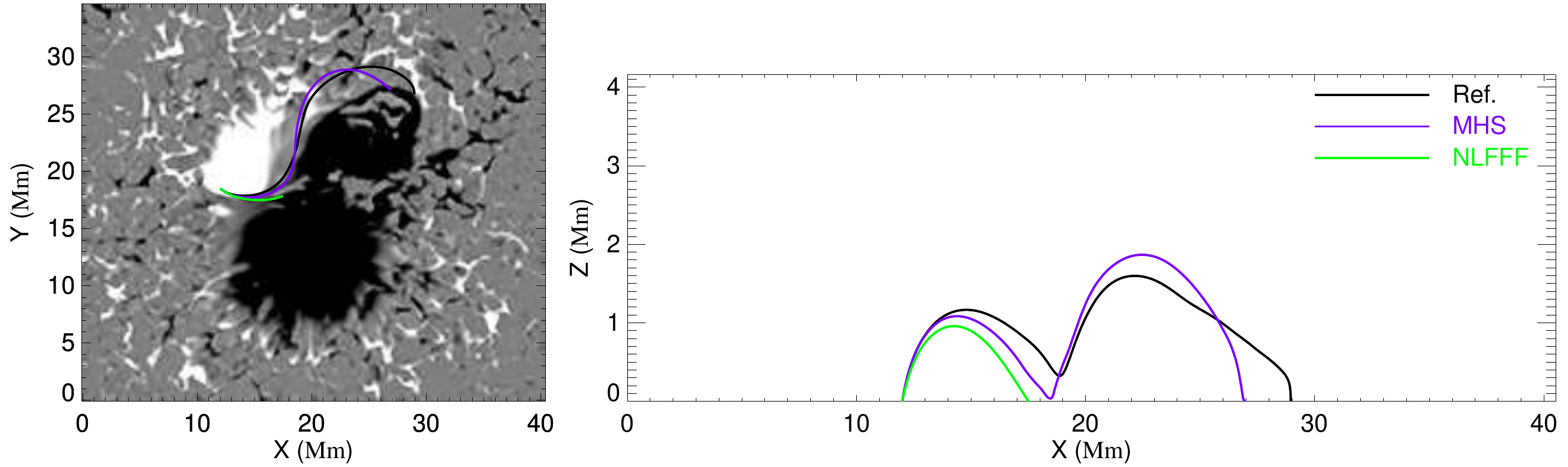}
\caption{Sample field lines of reference/MHS/NLFFF model from above (left) and from the south (right). The vertical scale of the right panel has been expanded.}
\label{fig:linedeviation}
\end{figure}

Fig.~\ref{fig:linecompare} shows the differences of the flux tube which is widely studied in the solar activity. The field lines of different models start from the same seeds where the vertical electric current is strong. The comparison denotes the MHS model is superior to the NLFFF model in reconstructing long twisted field lines. This leads to not only a different amount of twist, but also the connectivity of the magnetic field. As illustrated in Fig.~\ref{fig:linedeviation}, the left footpoint has a connection with another footpoint far away in the twin emerging spot in the reference and MHS model. In the NLFFF, however, it connects to the close side of the spot that exists all the time. The NLFFF line simply fails to bend up before it touch the photosphere, which results in large difference in connectivity. It is worth noting that the vector magnetogram used in NLFFF extrapolation is preprocessed to smooth and fulfill the force-free conditions. Without preprocessing the photospheric boundary data are inconsistent with the force-free assumption and consequently force-free computations do not converge (not show here). There have been some studies reported that the preprocessing improves the extrapolation considerably \citep{wis06,mds08}.

To quantify how well each extrapolation performed, we use 6 metrics as introduced by \cite{sdm06} and \cite{blw06} to compare vector field $\mathbf{b}$ (our results) and $\mathbf{B}$ (the reference MHD model):
\begin{enumerate}
\item[$\cdot$] vector correlation
\begin{equation}
C_{vec}=\displaystyle\sum_{i}\mathbf{B}_{i}\cdot \mathbf{b}_{i}/\left(\displaystyle\sum_{i}|\mathbf{B}_{i}|^2\displaystyle\sum_{i}|\mathbf{b}_{i}|^2\right)^{\frac{1}{2}},
\end{equation}
\item[$\cdot$] Cauchy-Schwarz inequality
\begin{equation}
C_{CS}=\frac{1}{N}\displaystyle\sum_{i}\frac{\mathbf{B}_{i}\cdot \mathbf{b}_{i}}{|\mathbf{B}_{i}||\mathbf{b}_{i}|},
\label{eq:ccs}
\end{equation}
\item[$\cdot$] normalized vector error
\begin{equation}
E_{N}=\displaystyle\sum_{i}|\mathbf{B}_{i}-\mathbf{b}_{i}|/\displaystyle\sum_{i}|\mathbf{B}_{i}|,
\end{equation}
\item[$\cdot$] mean vector error
\begin{equation}
E_{M}=\frac{1}{N}\displaystyle\sum_{i}\frac{|\mathbf{B}_{i}-\mathbf{b}_{i}|}{|\mathbf{B}_{i}|},
\label{eq:em}
\end{equation}
where N is the number of grid points in the computation box.
\item[$\cdot$] magnetic energy metric
\begin{equation}
E_{e}=\displaystyle\sum_{i}\frac{b_{i}^2}{2}/\displaystyle\sum_{i}\frac{B_{i}^2}{2},
\end{equation}
\item[$\cdot$] field line divergence (FLD) metric: for both the simulation and extrapolations, a score can be given to any point where the fied line is closed by the distance between the endpoints divided by the length of the field line. A single score can be assigned by the fraction of the area at the lower boundary that has an FLD less than 20\% (higher than 10\% used in \cite{blw06} and \cite{mds08}).
\end{enumerate}
The domain of comparison (within grid size of $210\times180\times64$) has the same field-of-view with Fig.~\ref{fig:magnetogram_linepattern} (b), extending upward to the height 4.1 Mm.

A quantitative evaluation of the two models is given in Tab.~\ref{tab:merit}. The MHS model scores better than NLFFF in all metrics. It is not clear that, based on the tiny difference of $C_{vec}$ and $C_{CS}$, MHS extrapolation is a significant improvement over the NLFFF extrapolation. However, for metric $1-E_m$ the MHS extrapolation scores 0.65 which is an improvement over the NLFFF's score of 0.57. It should be noted that $C_{vec}$ and $C_{CS}$ are sensitive to angle differences between the two vectors being compared, whereas $E_n$ and $E_m$ are sensitive to both angle and norm differences. The FLD is the most rigorous metric over the six we used. It also indicates that the MHS model performs better than the NLFFF model.

\begin{figure}[t]
\centering
\includegraphics[width=\hsize*2/3]{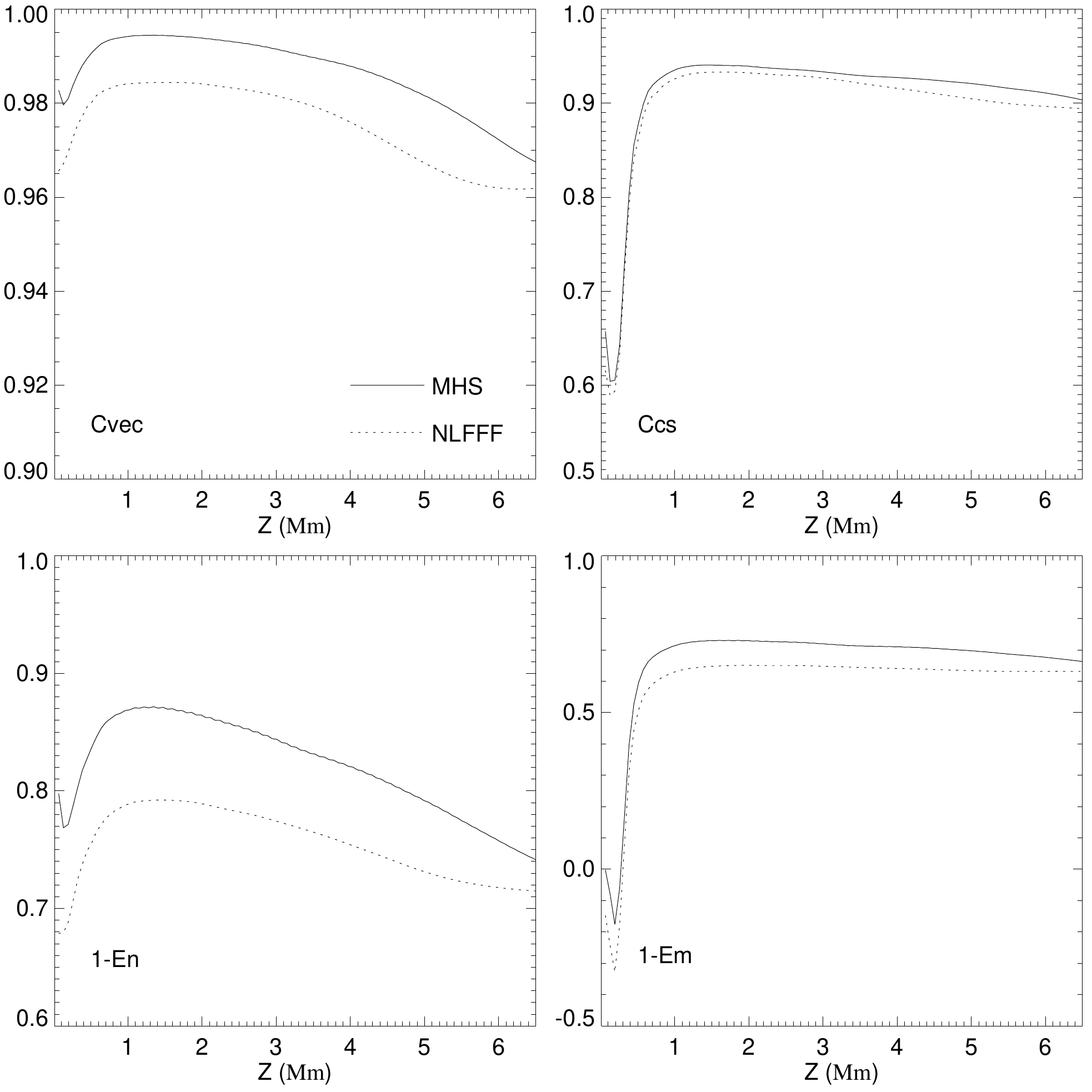}
\caption{Magnetic field metrics vary along height.}
\label{fig:metric}
\end{figure}

We also present the comparison of the first four metrics in every plane along the height (see Fig.~\ref{fig:metric}). Again, the MHS extrapolation scores better. It is worth noting that the weak field region can contribute significantly to the metrics $C_{CS}$ and $E_M$ (see Eqs.~\ref{eq:ccs} and \ref{eq:em}). That is the reason why, in the layers near the photosphere, the two metrics drop dramatically. The scores of two different extrapolations getting closer as the height increases, which means the decreasing influence of plasma effect with increasing height.

\begin{table}[b]
\caption{Merit for the two extrapolations applied to the vector magnetogram of the simulation. ``MHS$^a$'' represents the MHS extrapolation with all variables provided in the bottom boundary.}             
\label{tab:merit}      
\centering                          
\begin{tabular}{c c c c c c c}        
\hline\hline                 
\noalign{\smallskip}
Model & $C_{vec}$ & $C_{CS}$ & $1-E_{m}$ & $1-E_{n}$ & $E_{e}$ & $FLD$ \\    
\hline                        
\noalign{\smallskip}
   Ref.  & 1.00 & 1.00 & 1.00 & 1.00 & 1.00 & 1.00 \\      
   \noalign{\smallskip}
   NLFFF & 0.98 & 0.90 & 0.77 & 0.57 & 0.93 & 0.21 \\
   \noalign{\smallskip}
   MHS          & 0.99 & 0.91 & 0.85 & 0.65 & 0.95 & 0.44 \\
   \noalign{\smallskip}
   MHS$^{a}$    & 0.99 & 0.91 & 0.85 & 0.67 & 0.94 & 0.47 \\
   \noalign{\smallskip}
\hline                                   
\end{tabular}
\end{table}

\subsection{Magnetic twist and quasi-separatrix layers (QSLs) in the models}

\begin{figure}[t]
\centering
\includegraphics[width=\hsize*2/3]{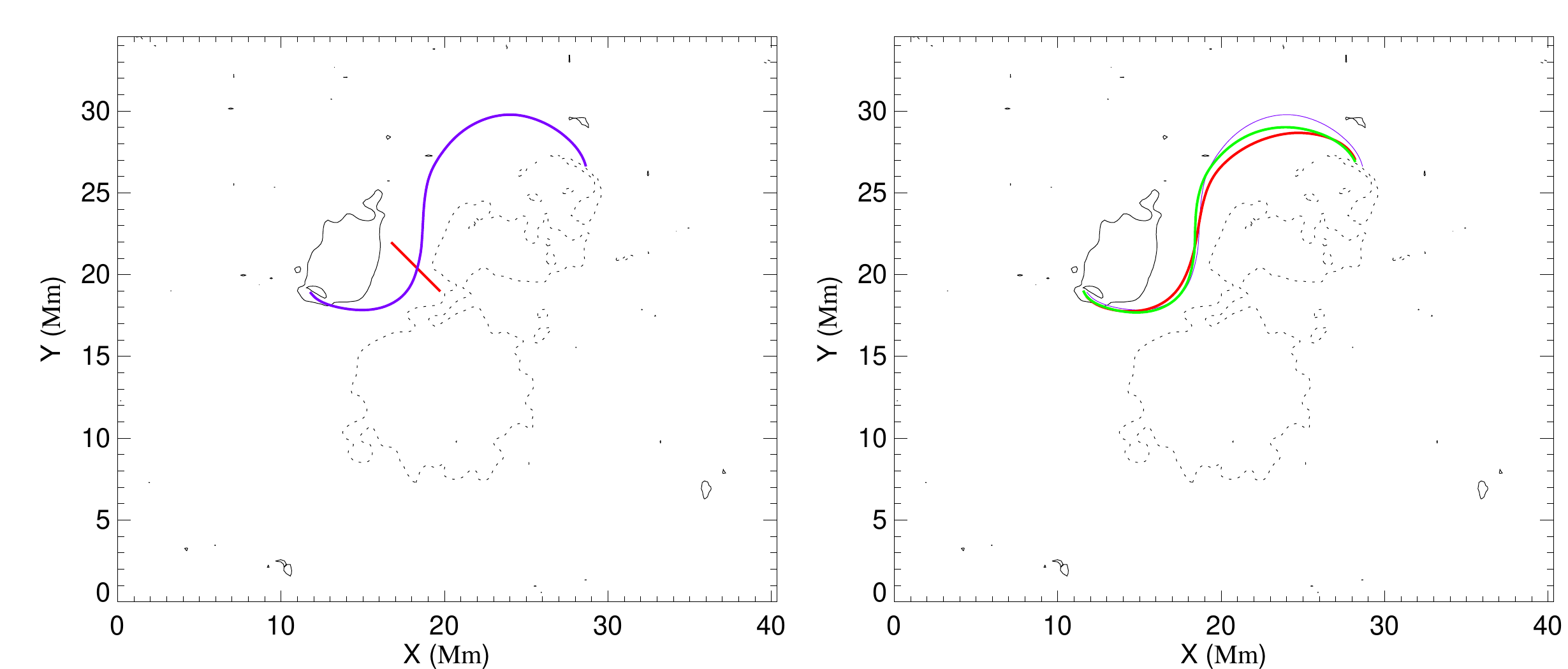}
\caption{Axis and surrounding field lines of the reference model. Left: Axis (purple) determined by the average $T_g$ of the surrounding field lines. Black/dotted contours denote positive/negative polarity. Right: Sample field lines of the flux rope.}
\label{fig:axis}
\end{figure}

Finally, we compare the magnetic twist numbers $T_w$, $T_g$ and squashing factor $Q_s$ of different models. $T_w$ and $T_g$ are defined by \cite{bp06} as the twist number of one magnetic field line about the neighboring line and the axis respectively. They are different in most situations. The self helicity which could be measured by local twist $T_w$ is negligible for a large number of flux tube \citep{dpb06}. In this study, $T_g$ is applicable to evaluate the twist since there exists only one flux rope with finite size. $T_w$ and $T_g$ are given by:
\begin{eqnarray}
T_w &=& \int_{s} \frac{\mu_{0}J_{\|}}{4\pi|\mathbf{B}|}ds=\int_{s} \frac{(\nabla \times \mathbf{B})\cdot \mathbf{B}}{4\pi B^{2}}ds,\\
T_g &=& \frac{1}{2\pi}\int_{s}\mathbf{T}(s)\cdot \mathbf{V}(s)\times \frac{d\mathbf{V}(s)}{ds}ds,\\
\end{eqnarray}
where $\mathbf{T}(s)$ is the unit tangent vector to the axis curve, $\mathbf{V}(s)$ is a unit vector normal to $\mathbf{T}(s)$ and points to the secondary curve, $J_{\|}$ is the parallel component of the electric current. The integration is carried out along the specific field line. QSLs are the generalized topological structures \citep{dhp96} which are the regions with high squashing factor $Q_s$. $Q_s$ is defined by mapping the field line \citep{thd02}:
\begin{eqnarray}
D_{12} & = & \left(
\begin{array}{cc}
\partial x_2/\partial x_1 \quad \partial x_2/\partial y_1\\
\partial y_2/\partial x_1 \quad \partial y_2/\partial y_1
\end{array}\right)
=  \left(
\begin{array}{cc}
a \qquad b \\
c \qquad d
\end{array}\right), \\
Q_s  &=&  \frac{a^2+b^2+c^2+d^2}{|B_n(x_1,y_1)/B_n(x_2,y_2)|},
\end{eqnarray}
where ($x_1$, $y_1$) and ($x_2$, $y_2$) are the two footpoints of one field line. Both indexes are important in studying the 3D magnetic structure. We used the code developed by \cite{lkt16} to calculate $T_w$ and $Q_s$. To compute $T_g$, we first have to locate the axis. We assume the axis as any one of magnetic field lines that penetrate the vertical square ($4\times4Mm$, thick black line from the top view showed in the left panel of Fig.~\ref{fig:axis}), a score can be assigned to this line based on the average $T_g$ of the surrounding lines. Then we define the axis as the field line whose score is the greatest (see the axis and sample surrounding field lines in Fig.~\ref{fig:axis}). For each model, the axis is determined by the same way. As a result, the average $T_g$ about the axes of the reference, MHS and NLFFF model is 0.95, 0.94 and 0.88, respectively. The twists of the field lines close to the axis are well reconstructed by the MHS model.

\begin{figure}[t]
\centering
\includegraphics[width=\hsize*1/2]{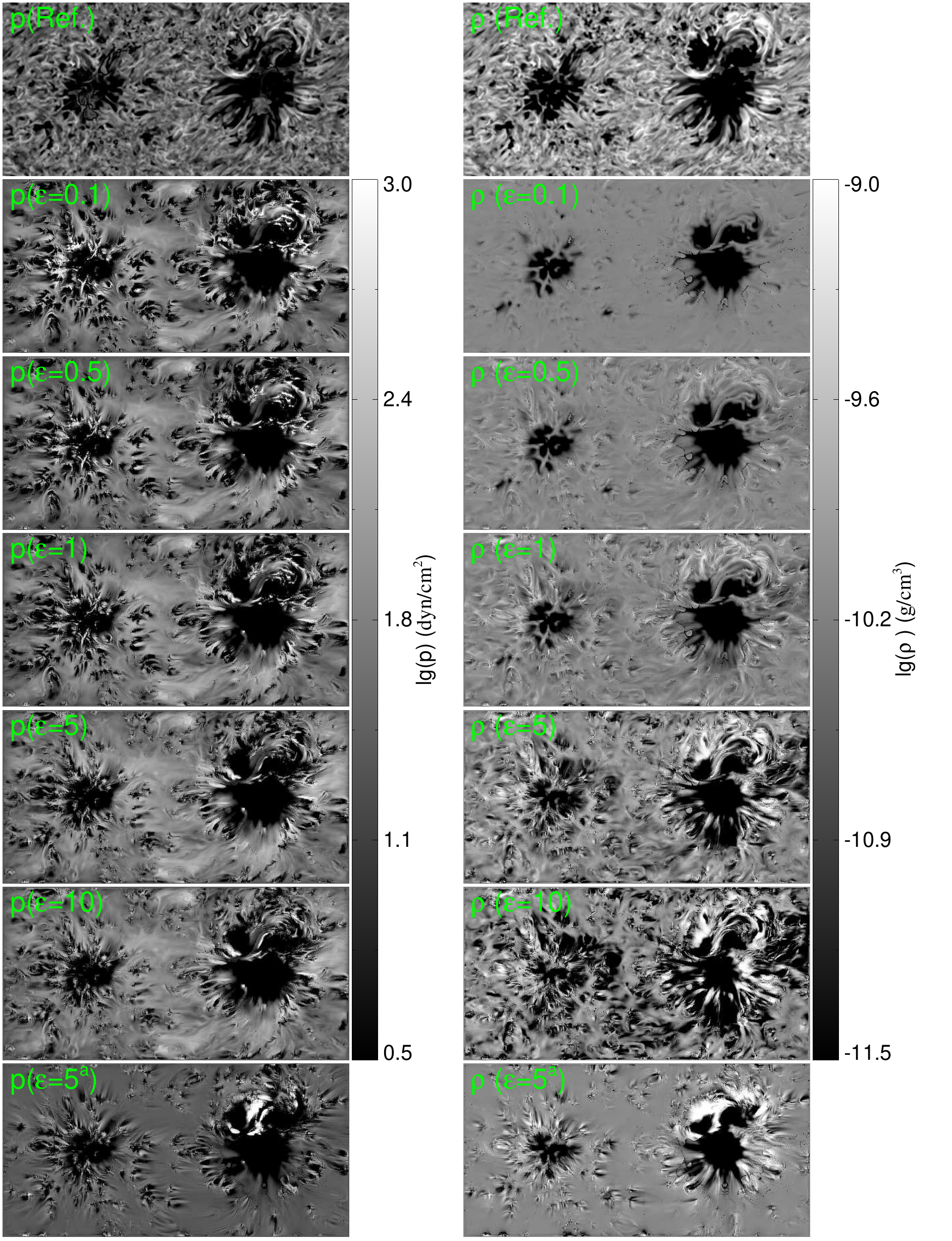}
\caption{Pressure and density at height 0.96 Mm with different $\epsilon$. $\epsilon=5^{a}$ represents the case with the potential field as the initial condition.}
\label{fig:pd_mu}
\end{figure}

\begin{figure}[t]
\centering
\includegraphics[width=\hsize*2/3]{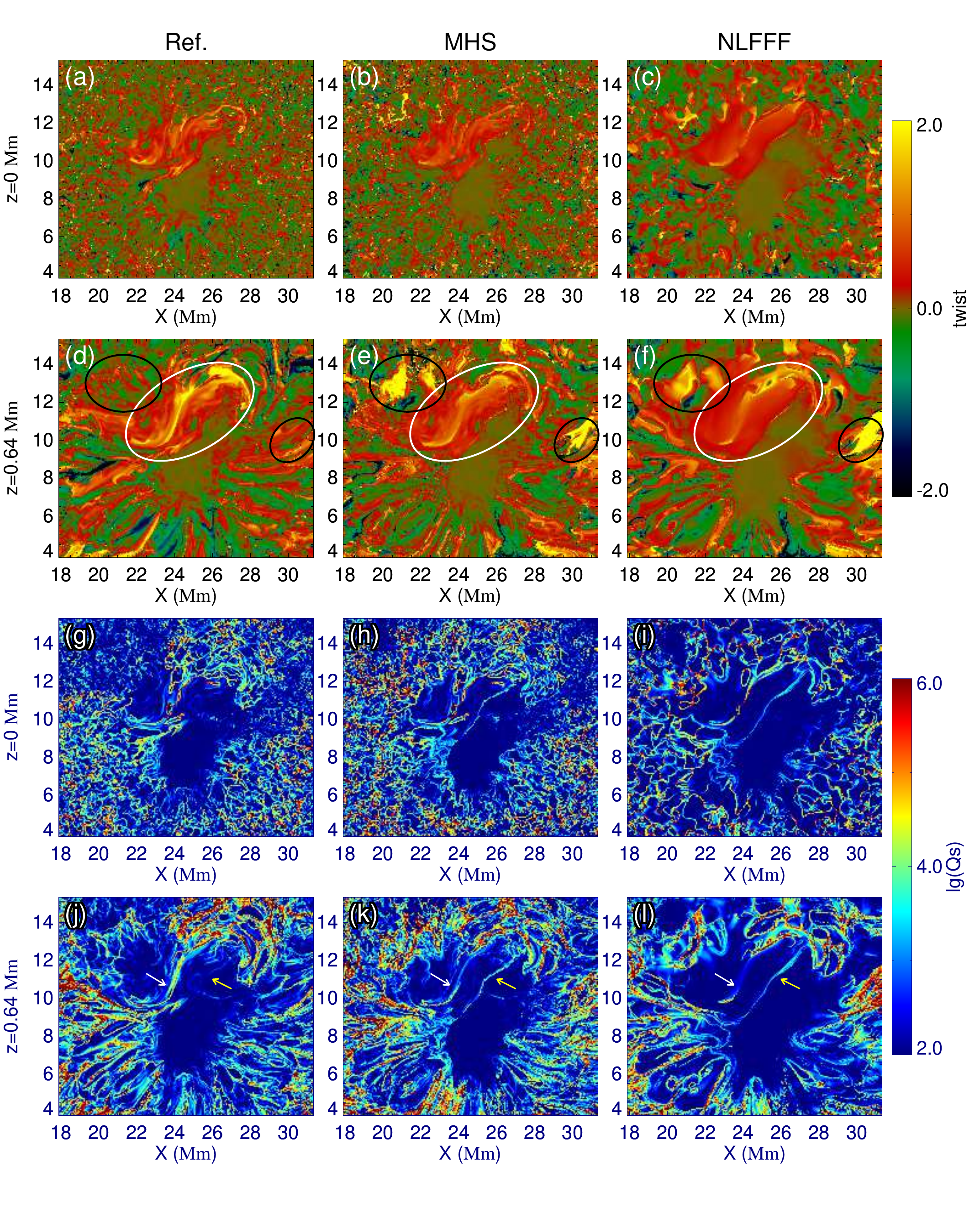}
\caption{$T_w$ and $Q_s$ comparisons at different levels according to the FOV of Fig.~\ref{fig:magnetogram_linepattern} (b).}
\label{fig:twq}
\end{figure}

The twist number $T_w$ is illustrated in Fig.~\ref{fig:twq} (a-f). The white ellipses represent the central area that is mainly recovered whereas the black ellipses show the mismatched areas. We can see the similar mismatch in the two extrapolations since the MHS model uses the NLFFF solution as the initial condition. If we look to some $T_w$ structures with small size, the MHS model includes more of them that exist in the reference model. That is because the MHS model uses the unpreprocessed magnetogram while the NLFFF model uses the preprocessed one. Without preprocessing, NLFFF does not converge because data are inconsistent with the force-free assumption. Preprocessing makes the data force-free consistent, but does naturally remove structures related to finite Lorentz-forces.

The QSLs represent the areas separate magnetic systems with different topology. We can see the QSLs pointed by the white arrow (Fig.~\ref{fig:twq}J) mainly divide the central area into two topological parts. However, there are additional QSLs in the two extrapolations, which divide the same area into three parts. So we should be very careful to use the QSLs result from extrapolation.

\begin{table}[b]
\caption{Metrics for different $\epsilon$. $\epsilon=5^{a}$ represents the case with the potential field as the initial condition.}             
\label{tab:epsilon}      
\centering                          
\begin{tabular}{c c c c c c c}        
\hline\hline                 
\noalign{\smallskip}
$\epsilon$ & $C_{vec}$ & $C_{CS}$ & $1-E_{m}$ & $1-E_{n}$ & $L$ & \\    
\hline                        
\noalign{\smallskip}
0.1 & 0.9890 & 0.9025 & 0.8340 & 0.6341 & 5.2 \\      
\noalign{\smallskip}
0.5 & 0.9899 & 0.9039 & 0.8398 & 0.6401 & 3.4 \\
\noalign{\smallskip}
1   & 0.9903 & 0.9054 & 0.8420 & 0.6430 & 3.0 \\
\noalign{\smallskip}
5   & 0.9908 & 0.9068 & 0.8453 & 0.6481 & 2.7 \\
\noalign{\smallskip}
10  & 0.9907 & 0.9060 & 0.8451 & 0.6472 & 2.5 \\
\noalign{\smallskip} 
$5^{a}$  & 0.9720 & 0.8597 & 0.7624 & 0.5357 & 8.4 \\
\noalign{\smallskip} 
\hline                                 
\end{tabular}
\end{table}

\section{Discussion}\label{sec:discussion}

The inability to generate MHS model that well recover the reference model is rather disappointing. This cause us to exam the new extrapolation to find out some factors that impact out ability to produce a robust model.

\subsection{Choice of free parameters and initial conditions}

Three parameters $\mu_{B}$, $\mu_{p}$ and $\mu_{\rho}$ control the step length when the Eq.~\ref{eq:L} is minimizing. $\mu_{B}$ is computed every step by the line search method. We assume $\mu_{p}=\mu_{B}$ because $Q$ has the same dimension as $B$. We further assume $\mu_{\rho}=\epsilon \mu_{B}$. Table \ref{tab:epsilon} shows the case study with $\epsilon$ varied between 0.1 and 10. We find the functional L decreases with increasing $\epsilon$. However, the most accurate magnetic field generated by the code when $\epsilon=5$. Fig.~\ref{fig:pd_mu} shows how the free parameter $\epsilon$ affect the pressure and density result. The density is much more affected than the pressure by $\epsilon$. A larger $\epsilon$ leads to a stronger contrast of the density because of the higher weight introduced by $\epsilon$. From the above cases study, we choose $\epsilon=5$ as the optimized value for the extrapolation.


The choice of the initial condition of magnetic field has significant influence on the results \citep{w04,sdm06,zw18}. We run an additional test of $\epsilon=5$ with potential field as the initial condition. The test is named as $5^{a}$ in table \ref{tab:epsilon} and Fig.~\ref{fig:pd_mu}. It shows that the MHS extrapolation with the initial potential field has lower scores of the magnetic field evaluation and less accurate plasma distribution.

\subsection{All variables in the bottom boundary provided}

\begin{figure}[h]
\centering
\includegraphics[width=\hsize*2/3]{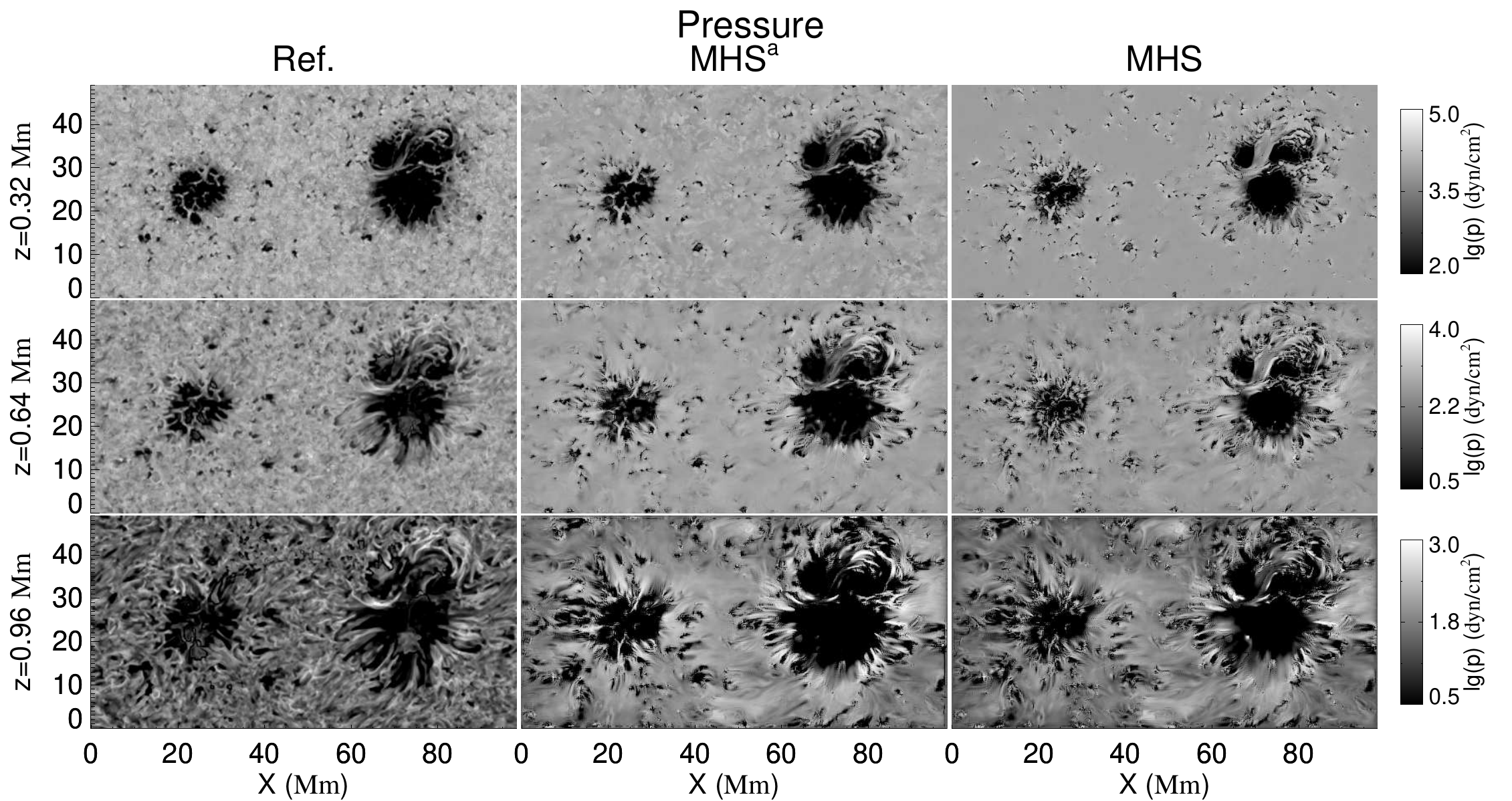}
\caption{Pressure comparison at different levels.}
\label{fig:p_btm}
\end{figure}

\begin{figure}[b]
\centering
\includegraphics[width=\hsize*2/3]{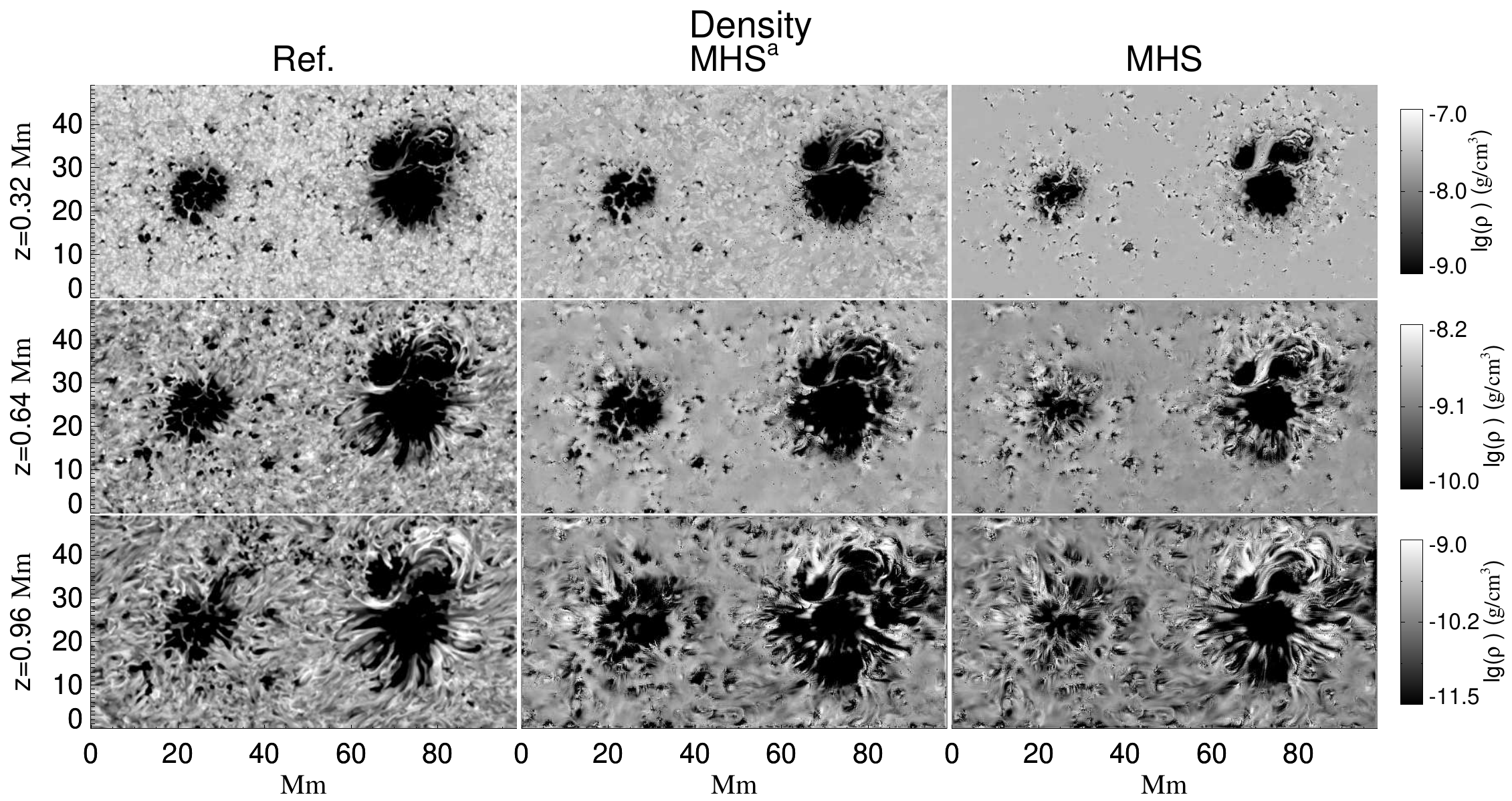}
\caption{Density comparison at different levels.}
\label{fig:d_btm}
\end{figure}

Table \ref{tab:merit} also compares the magnetic field of reference model with the result for case ``MHS$^{a}$'' for which the pressure and density in the bottom boundary is provided. We see limited increase of the metrics for the magnetic field comparison. Fig.~\ref{fig:p_btm} and \ref{fig:d_btm} show the plasma distribution in the extremely low height ($\le 0.32Mm$) benefit from the use of the accurate bottom boundary condition. Above 1 Mm, however, the plasma result shows no improvement.

The increasing dynamics in the higher atmosphere make the result insensitive to the choice of the plasma boundary condition in the bottom. Nevertheless, a simplest boundary condition with the plasma uniformly distributed is not able to yield good result. So the plasma boundary described in Setction 3.1 is still a good choice.



\section{Conclusions}\label{sec:conclusion}

In this work, we applied the MHS extrapolation code to model an active region which is a snapshot of the flare simulation. The magnetic field, plasma pressure and density are computed consistently by the model.

The MHS model is able to reconstruct the main structures of pressure and density in the photosphere and lower choromosphere (below 1 Mm). As we did not use the temperature data to constrain the plasma, the plasma solution above 1 Mm cannot be trusted. The deviation is getting larger as the height increases.

Generally, the magnetic field solution is improved compared with the NLFFF not only on the magnitude and direction, but also on the magnetic connectivity. As to the flux tube which is of vital importance in the eruptive events, the twists of the field lines are well constructed. However, some twist structures derived outside the active region are incorrect. It is worth noting that even right inside the active region, the extrapolated QSLs could be unreal. The $Q_s$ value is computed by tracing the field lines from which the high nonlinearity is introduced. A tiny deviation showed in Fig.~\ref{fig:linedeviation} near the bald patch area results in large difference of connectivity.

Paper I developed the MHS extrapolation and tested the model using a semi-analytic MHS solution. In this work, we present a more challenging test of our model with a radiative MHD simulation of a solar flare. An application to the IMaX magnetogram \citep{mda11} on board SUNRISE balloon-borne solar observatory \citep{bgs11,bss11} during its second flight \citep{srb17} is planned.

\begin{appendix} 
\section{Variable definitions} \label{sec:variables}
The variables in Eqs.~ \ref{eq:Bn+1}-\ref{eq:Rn+1} are defined as
\begin{eqnarray}
\delta L_{\mathbf{B}}&=&\mathbf{\tilde{F}}_{a}+\mathbf{\tilde{F}}_{b},\\
\delta L_{Q}&=&2[\lambda(\omega_{a}\Omega_{a}^{2}+\omega_{b}\Omega_{b}^{2})-\nabla \cdot (\mathbf{\omega_{a}\tilde{\Omega}_{a}})]Q,\\
\delta L_{R}&=&2\omega_{a}R\mathbf{\tilde{\Omega}_{a}}\cdot \mathbf{\hat{z}},\\
\\
\mathbf{\tilde{F}}_{a}&=&\omega_{a}\mathbf{F}_{a}+(\mathbf{\tilde{\Omega}}_{a}\times \mathbf{B})\times \nabla\omega_{a},\\
\mathbf{\tilde{F}}_{b}&=&\omega_{b}\mathbf{F}_{b}+(\mathbf{\Omega}_{b}\cdot  \mathbf{B}) \nabla\omega_{b},\\
\mathbf{F}_{a}&=&\nabla\times(\mathbf{\tilde{\Omega}}_{a}\times\mathbf{B})-\mathbf{\tilde{\Omega}}_{a}\times(\nabla\times\mathbf{B})+(1-2\lambda)\Omega_{a}^{2}\mathbf{B},\\
\mathbf{F}_{b}&=&\nabla(\mathbf{\Omega}_{b}\cdot\mathbf{B})-\mathbf{\Omega}_{b}(\nabla\cdot\mathbf{B})+\Omega_{b}^{2}\mathbf{B},\\
\mathbf{\tilde{\Omega}}_{a}&=&\lambda\mathbf{\Omega}_{a},\\
\lambda&=&\frac{B^{2}}{B^{2}+Q^{2}}
\end{eqnarray}
where $\mathbf{\hat{n}}$ is the inward unit vector on the surface S.
\end{appendix}

\begin{acknowledgements}
We appreciate the constructive comments from the anonymous referee. This work was supported by DFG-grant WI 3211/4-1.
\end{acknowledgements}

%
   \bibliographystyle{aa} 
   \bibliography{aa} 

\begin{thebibliography}{43}
\expandafter\ifx\csname natexlab\endcsname\relax\def\natexlab#1{#1}\fi

\bibitem[{{Aulanier} {et~al.}(1999){Aulanier}, {D{\'e}moulin}, {Mein}, {van
  Driel-Gesztelyi}, {Mein}, \& {Schmieder}}]{adm99}
{Aulanier}, G., {D{\'e}moulin}, P., {Mein}, N., {et~al.} 1999, \aap, 342, 867

\bibitem[{{Aulanier} {et~al.}(1998){Aulanier}, {D{\'e}moulin}, {Schmieder},
  {Fang}, \& {Tang}}]{ads98}
{Aulanier}, G., {D{\'e}moulin}, P., {Schmieder}, B., {Fang}, C., \& {Tang},
  Y.~H. 1998, \solphys, 183, 369

\bibitem[{{Barnes} {et~al.}(2006){Barnes}, {Leka}, \& {Wheatland}}]{blw06}
{Barnes}, G., {Leka}, K.~D., \& {Wheatland}, M.~S. 2006, \apj, 641, 1188

\bibitem[{{Barthol} {et~al.}(2011){Barthol}, {Gandorfer}, {Solanki},
  {Sch{\"u}ssler}, {Chares}, {Curdt}, {Deutsch}, {Feller}, {Germerott}, \&
  {Grauf}}]{bgs11}
{Barthol}, P., {Gandorfer}, A., {Solanki}, S.~K., {et~al.} 2011, \solphys, 268,
  1

\bibitem[{{Berger} \& {Prior}(2006)}]{bp06}
{Berger}, M.~A. \& {Prior}, C. 2006, Journal of Physics A Mathematical General,
  39, 8321

\bibitem[{{Berkefeld} {et~al.}(2011){Berkefeld}, {Schmidt}, {Soltau}, {Bell},
  {Doerr}, {Feger}, {Friedlein}, {Gerber}, {Heidecke}, \& {Kentischer}}]{bss11}
{Berkefeld}, T., {Schmidt}, W., {Soltau}, D., {et~al.} 2011, \solphys, 268, 103

\bibitem[{{Cheung} {et~al.}(2019){Cheung}, {Rempel}, {Chintzoglou}, {Chen},
  {Testa}, {Mart{\'\i}nez-Sykora}, {Sainz Dalda}, {DeRosa}, {Malanushenko}, \&
  {Hansteen}}]{crc19}
{Cheung}, M.~C.~M., {Rempel}, M., {Chintzoglou}, G., {et~al.} 2019, Nature
  Astronomy, 3, 160

\bibitem[{{Demoulin} {et~al.}(1996){Demoulin}, {Henoux}, {Priest}, \& {Mand
  rini}}]{dhp96}
{Demoulin}, P., {Henoux}, J.~C., {Priest}, E.~R., \& {Mand rini}, C.~H. 1996,
  \aap, 308, 643

\bibitem[{{Demoulin} {et~al.}(2006){Demoulin}, {Pariat}, \& {Berger}}]{dpb06}
{Demoulin}, P., {Pariat}, E., \& {Berger}, M.~A. 2006, \solphys, 233, 3

\bibitem[{{Gary}(2001)}]{g01}
{Gary}, G.~A. 2001, \solphys, 203, 71

\bibitem[{{Gilchrist} {et~al.}(2016){Gilchrist}, {Braun}, \& {Barnes}}]{gbb16}
{Gilchrist}, S.~A., {Braun}, D.~C., \& {Barnes}, G. 2016, \solphys, 291, 3583

\bibitem[{{Gilchrist} \& {Wheatland}(2013)}]{gw13}
{Gilchrist}, S.~A. \& {Wheatland}, M.~S. 2013, \solphys, 282, 283

\bibitem[{{Guo} {et~al.}(2017){Guo}, {Cheng}, \& {Ding}}]{gcd17}
{Guo}, Y., {Cheng}, X., \& {Ding}, M. 2017, Science in China Earth Sciences,
  60, 1408

\bibitem[{{Hu} \& {Dasgupta}(2006)}]{hd06}
{Hu}, Q. \& {Dasgupta}, B. 2006, \grl, 33, L15106

\bibitem[{{Hu} \& {Dasgupta}(2008)}]{hd08}
{Hu}, Q. \& {Dasgupta}, B. 2008, \solphys, 247, 87

\bibitem[{{Jiao} {et~al.}(1997){Jiao}, {McClymont}, \& {Mikic}}]{jmm97}
{Jiao}, L., {McClymont}, A.~N., \& {Mikic}, Z. 1997, \solphys, 174, 311

\bibitem[{{Liu} {et~al.}(2016){Liu}, {Kliem}, {Titov}, {Chen}, {Wang}, {Wang},
  {Liu}, {Xu}, \& {Wiegelmann}}]{lkt16}
{Liu}, R., {Kliem}, B., {Titov}, V.~S., {et~al.} 2016, \apj, 818, 148

\bibitem[{{Low}(1985)}]{l85}
{Low}, B.~C. 1985, \apj, 293, 31

\bibitem[{{Low}(1991)}]{l91}
{Low}, B.~C. 1991, \apj, 370, 427

\bibitem[{{Low}(1992)}]{l92}
{Low}, B.~C. 1992, \apj, 399, 300

\bibitem[{{Mart{\'\i}nez Pillet} {et~al.}(2011){Mart{\'\i}nez Pillet}, {Del
  Toro Iniesta}, {{\'A}lvarez-Herrero}, {Domingo}, {Bonet}, {Gonz{\'a}lez
  Fern{\'a}ndez}, {L{\'o}pez Jim{\'e}nez}, {Pastor}, {Gasent Blesa}, \&
  {Mellado}}]{mda11}
{Mart{\'\i}nez Pillet}, V., {Del Toro Iniesta}, J.~C., {{\'A}lvarez-Herrero},
  A., {et~al.} 2011, \solphys, 268, 57

\bibitem[{{McClymont} \& {Mikic}(1994)}]{mm94}
{McClymont}, A.~N. \& {Mikic}, Z. 1994, \apj, 422, 899

\bibitem[{{Metcalf} {et~al.}(2008){Metcalf}, {De Rosa}, {Schrijver}, {Barnes},
  {van Ballegooijen}, {Wiegelmann}, {Wheatland}, {Valori}, \&
  {McTtiernan}}]{mds08}
{Metcalf}, T.~R., {De Rosa}, M.~L., {Schrijver}, C.~J., {et~al.} 2008,
  \solphys, 247, 269

\bibitem[{{Metcalf} {et~al.}(1995){Metcalf}, {Jiao}, {McClymont}, {Canfield},
  \& {Uitenbroek}}]{mjm95}
{Metcalf}, T.~R., {Jiao}, L., {McClymont}, A.~N., {Canfield}, R.~C., \&
  {Uitenbroek}, H. 1995, \apj, 439, 474

\bibitem[{{Neukirch} \& {Rast{\"a}tter}(1999)}]{nr99}
{Neukirch}, T. \& {Rast{\"a}tter}, L. 1999, \aap, 348, 1000

\bibitem[{{Rempel}(2017)}]{r17}
{Rempel}, M. 2017, \apj, 834, 10

\bibitem[{{Schrijver} {et~al.}(2006){Schrijver}, {De Rosa}, {Metcalf}, {Liu},
  {McTiernan}, {R{\'e}gnier}, {Valori}, {Wheatland}, \& {Wiegelmann}}]{sdm06}
{Schrijver}, C.~J., {De Rosa}, M.~L., {Metcalf}, T.~R., {et~al.} 2006,
  \solphys, 235, 161

\bibitem[{{Solanki} {et~al.}(2017){Solanki}, {Riethm{\"u}ller}, {Barthol},
  {Danilovic}, {Deutsch}, {Doerr}, {Feller}, {Gandorfer}, {Germerott}, \&
  {Gizon}}]{srb17}
{Solanki}, S.~K., {Riethm{\"u}ller}, T.~L., {Barthol}, P., {et~al.} 2017,
  \apjs, 229, 2

\bibitem[{{Titov} {et~al.}(2002){Titov}, {Hornig}, \& {D{\'e}moulin}}]{thd02}
{Titov}, V.~S., {Hornig}, G., \& {D{\'e}moulin}, P. 2002, Journal of
  Geophysical Research (Space Physics), 107, 1164

\bibitem[{{V{\"o}gler} {et~al.}(2005){V{\"o}gler}, {Shelyag}, {Sch{\"u}ssler},
  {Cattaneo}, {Emonet}, \& {Linde}}]{vss05}
{V{\"o}gler}, A., {Shelyag}, S., {Sch{\"u}ssler}, M., {et~al.} 2005, \aap, 429,
  335

\bibitem[{{Wiegelmann}(2004)}]{w04}
{Wiegelmann}, T. 2004, \solphys, 219, 87

\bibitem[{{Wiegelmann} \& {Inhester}(2003)}]{wi03}
{Wiegelmann}, T. \& {Inhester}, B. 2003, \solphys, 214, 287

\bibitem[{{Wiegelmann} {et~al.}(2006){Wiegelmann}, {Inhester}, \&
  {Sakurai}}]{wis06}
{Wiegelmann}, T., {Inhester}, B., \& {Sakurai}, T. 2006, \solphys, 233, 215

\bibitem[{{Wiegelmann} \& {Neukirch}(2006)}]{wn06}
{Wiegelmann}, T. \& {Neukirch}, T. 2006, \aap, 457, 1053

\bibitem[{{Wiegelmann} {et~al.}(2017){Wiegelmann}, {Neukirch}, {Nickeler},
  {Solanki}, {Barthol}, {Gandorfer}, {Gizon}, {Hirzberger}, {Riethm{\"u}ller},
  \& {van Noort}}]{wnn17}
{Wiegelmann}, T., {Neukirch}, T., {Nickeler}, D.~H., {et~al.} 2017, \apjs, 229,
  18

\bibitem[{{Wiegelmann} {et~al.}(2015){Wiegelmann}, {Neukirch}, {Nickeler},
  {Solanki}, {Mart{\'\i}nez Pillet}, \& {Borrero}}]{wnn15}
{Wiegelmann}, T., {Neukirch}, T., {Nickeler}, D.~H., {et~al.} 2015, \apj, 815,
  10

\bibitem[{{Wiegelmann} {et~al.}(2007){Wiegelmann}, {Neukirch}, {Ruan}, \&
  {Inhester}}]{wnr07}
{Wiegelmann}, T., {Neukirch}, T., {Ruan}, P., \& {Inhester}, B. 2007, \aap,
  475, 701

\bibitem[{{Wiegelmann} \& {Sakurai}(2012)}]{ws12}
{Wiegelmann}, T. \& {Sakurai}, T. 2012, Living Reviews in Solar Physics, 9, 5

\bibitem[{{Wiegelmann} {et~al.}(2012){Wiegelmann}, {Thalmann}, {Inhester},
  {Tadesse}, {Sun}, \& {Hoeksema}}]{wti12}
{Wiegelmann}, T., {Thalmann}, J.~K., {Inhester}, B., {et~al.} 2012, \solphys,
  281, 37

\bibitem[{{Zhu} {et~al.}(2017){Zhu}, {Wang}, {Cheng}, \& {Huang}}]{zwc17}
{Zhu}, X., {Wang}, H., {Cheng}, X., \& {Huang}, C. 2017, \apjl, 844, L20

\bibitem[{{Zhu} {et~al.}(2016){Zhu}, {Wang}, {Du}, \& {He}}]{zwd16}
{Zhu}, X., {Wang}, H., {Du}, Z., \& {He}, H. 2016, \apj, 826, 51

\bibitem[{{Zhu} \& {Wiegelmann}(2018)}]{zw18}
{Zhu}, X. \& {Wiegelmann}, T. 2018, \apj, 866, 130

\bibitem[{{Zhu} {et~al.}(2013){Zhu}, {Wang}, {Du}, \& {Fan}}]{zwd13}
{Zhu}, X.~S., {Wang}, H.~N., {Du}, Z.~L., \& {Fan}, Y.~L. 2013, \apj, 768, 119

\end{thebibliography}
%

\end{document}